\documentclass[twocolumn,showpacs,preprintnumbers,amsmath,amssymb]{revtex4}
\usepackage{graphicx}% Include figure files
\usepackage{dcolumn}% Align table columns on decimal point
\usepackage{bm}% bold math  
\begin{document}
%\preprint{APS/123-QED} 

\title{Hydrogen adsorption on boron doped graphene: an {\it ab initio} study}

\author{R. H.  Miwa$^1$, T. B. Martins$^2$, and A.  Fazzio$^2$}

\affiliation{$^1$Instituto de  F\'{\i}sica, Universidade  Federal de
Uberl\^andia,  Caixa Postal  593,  38400-902, Uberl\^andia,  MG, Brazil.}

\affiliation{$^2$Instituto de F\'{\i}sica, Universidade de S\~ao Paulo, Caixa Postal 66318, 05315-970, S\~ao Paulo, SP, Brazil.}
 
\date\today

\begin{abstract} 
  
The electronic  and structural properties of (i)  boron doped graphene
sheets, and  (ii) the chemisorption  processes of hydrogen  adatoms on
the boron doped graphene sheets  have been examined by {\it ab initio}
total  energy  calculations.   In  (i)  we find  that  the  structural
deformations are very localized around the boron substitutional sites,
and there is an increase of  the electronic density of states near the
Fermi level.  Our simulated  STM images, for occupied states, indicate
the   formation  of  a   bright  (triangular)   spots  lying   on  the
substitutional  boron  (center)  and  nearest neighbor  carbon  (edge)
sites.   Those STM  pictures are  attributed  to the  increase of  the
density of states within an  energy interval of 0.5~eV below the Fermi
level. For boron concentration of  $\sim$2.4\%, we find that two boron
atoms  lying  on  the  opposite  sites  of  the  same  hexagonal  ring
($B1$-$B2$  configuration) represents  the  energetically most  stable
configuration,  which   is  in  contrast   with  previous  theoretical
findings.    Having   determined   the   energetically   most   stable
configuration for  substitutional boron  atoms on graphene  sheets, we
next considered the  hydrogen adsorption process as a  function of the
boron concentration,  (ii).  Our calculated  binding energies indicate
that  the  C--H bonds  are  strengthened  nearby boron  substitutional
sites.  Indeed, the binding energy of hydrogen adatoms forming a dimer
like structure, aside $B1$-$B2$, is  higher than the binding energy of
an  isolated H$_2$  molecule.  Since  the  formation of  H dimer  like
structure may represents the  initial stage of the hydrogen clustering
process  on graphene  sheets, we  can infer  that the  formation  of H
clusters are quite likely not only on the clean graphene sheets, which
is  in  consonance with  previous  studies  [Phys.   Lett.  {\bf  97},
186102],  but  also on  the  $B1$-$B2$  boron  doped graphene  sheets.
However,  for low  concentration of  boron atoms,  the formation  of H
dimer   structures  is  not   expected  to   occur  nearby   a  single
substitutional boron site.   In this way, the formation  (or not) of H
clusters  on graphene  sheets can  be  tuned by  the concentration  of
substitutional boron atoms.

\end{abstract} 


\maketitle

\section{Introduction}

Two  dimensional  crystals  of  $sp^2$  bonded  carbon  atoms,  called
graphene, exhibit quite different electronic and structural properties
compared with their counterpart, {\it viz.}: graphite (graphene sheets
stacked in  ABABAB...  arrangement),  and carbon nanotubes  (rolled up
graphene sheets forming nanometer wide cylinders). Indeed, in a recent
experimental study,  Novoselov et al.~\cite{novoselovNat2005} observed
massless two dimensional (2D) Dirac fermions in free standing graphene
sheets  with  high  crystal  quality.   In  this  case,  the  electron
scattering processes are suppressed, and thus, the charge carriers can
propagate freely  on the graphene surface.   Meanwhile, very recently,
Meyer  at al.   observed microscopic  corrugations on  a  single layer
suspended graphene sheet, due to the carbon atoms displaced out of the
graphene   plane.    Such   an   experimental  finding   explain   the
thermodynamic stability of those isolated ``two dimensional'' systems,
however, ``further experimental and  theoretical studies are needed to
clarify  the  detailed mechanism  of  the  corrugations in  graphene''
~\cite{meyerNat2007}.

On  graphene sheets,  the  $sp^2$ orbitals,  forming $\sigma$  bonding
states  parallel to  the graphene  surface, connect  the  carbon atoms
arranged  in   a  hexagonal   lattice.   While  the   $p_z$  orbitals,
perpendicular to  the graphene sheet,  give rise to  delocalized $\pi$
bonding  and   $\pi^\ast$  anti-bonding  states.   The   most  of  the
electronic properties  and the chemical reactivity  of graphene sheets
are ruled by  those $\pi$ and $\pi^\ast$ orbitals.   For instance, the
already  mentioned  massless  2D  Dirac  fermions,  and  the  recently
observed control  of the electronic properties in  thin films composed
by  bilayer graphene  sheets~\cite{ohtaScience2006}.  Focusing  on the
chemical properties,  graphene has been considered as  a candidate for
hydrogen                                                        storage
media~\cite{schlapbachNature2001,patchkovskiiPNAS2006}.        Scanning
tunneling microscope  (STM) pictures, of hydrogen  adatoms on graphene
surfaces,  indicate a  local  electronic charge  enhancement and  long
range electronic perturbation on the  graphene sheet, giving rise to a
local      $(\sqrt       3\times      \sqrt      3)R30^o$      surface
periodicity~\cite{ruffieuxPRL2000}.  Hydrogen  on graphene surface has
been  the  subject  of   intense  studies,  not  only  addressing  the
realization of  fuel cells based  upon hydrogen storage, but  also the
formation of H$_2$  molecules in the interstellar medium  (which is an
important problem in astrophysics).

The  energetic  stability and  the  equilibrium  geometry of  hydrogen
adatoms on  graphene surfaces were  investigated in detail by  Sha and
Jackson~\cite{shaSUSC2002}  and Duplock  et al.~\cite{duplockPRL2004}.
They find  binding energies of 0.67 and  0.76~eV/H atom, respectively,
for a single hydrogen adatom chemisorbed on the graphene sheet. Having
characterized  the  energetic and  structural  properties of  hydrogen
adatoms on graphene, further experimental/theoretical studies examined
the formation of H$_2$  molecules through H--H recombination processes
on  the graphene  surface.   For instance,  the Eley--Rideal  reaction
involving  free  hydrogen  atoms  (gas  phase)  and  hydrogen  adatoms
chemisorbed                on               the               graphene
surface~\cite{shaSUSC2002,shaJChemPhys2002}.   On the other  hand, the
H--H  recombination  through   hydrogen  diffusion,  on  the  graphene
surface, represents  another reaction  path to obtain  H$_2$ molecules
from         the          chemisorbed         hydrogen         adatoms
\cite{ferroChemPhysLett2003,hornekaerPRL2006-ii,baoucheTheJChemPhys2006}.
Based upon those H--H reaction processes, recent experimental studies,
supported by  {\it ab initio} calculations, examined  the formation of
hydrogen clusters  on graphene surfaces~\cite{hornekaerPRL2006}. Their
findings  suggest the  formation  of  H dimer  like  structures as  an
initial  stage of the  observed hydrogen  clusters chemisorbed  on the
graphene surface.   The calculated binding energy  of hydrogen adatoms
forming a H dimer structure is higher compared with the binding energy
of two isolated H monomers.

The binding energy of hydrogen adatoms chemisorbed on graphene sheets,
as  well as  the formation  of H  clusters, can  be tuned  by suitable
doping processes.  For instance,  the presence of substitutional boron
atoms on graphene improve the  hydrogen adsorption on the carbon atoms
neighboring             the            boron            substitutional
sites~\cite{ferroJChemPhys2003,zhuJPhysChemB2006}.    However,   those
calculated  hydrogen   binding  energy  results,  as   a  function  of
substitutional  boron concentration,  are quite  contradictory.  Based
upon Raman  spectroscopy measurements, Endo  et al.~\cite{endoPRB1998}
characterized the  equilibrium geometry of  substitutional boron atoms
on  graphite.   For  boron  concentration  of  $\sim$2.7\%  they  find
homogeneous distribution of substitutional boron atoms.  Meanwhile, for
higher  concentration of  substitutional  boron atoms,  about 17\%,  a
detailed     experimental    study,     performed    by     Hach    et
al.~\cite{hachCarbon1999},   indicates   the   formation   of   $C_6B$
configuration. In the $C_6B$ structure, the substitutional boron atoms
occupy  the  opposite  sites  of  a carbon  hexagonal  ring.   Further
experimental/theoretical studies have been  done aiming to improve our
knowledge       related       to       boron      doped       graphite
systems~\cite{endoCarbon1999,endoJAP2001,martinezPRL2007}.

In  the  present paper  we  report an  {\it  ab  initio} total  energy
investigation of  boron doped  graphene sheets, and  the chemisorption
processes of  hydrogen adatoms as  a function of  boron concentration.
The presence  of substitutional  boron atoms increases  the electronic
density of  states near  the Fermi level.   Our calculated  STM images
indicate  the formation of  bright spots  on the  boron substitutional
sites,  which is  in agreement  with the  experimentally  obtained STM
pictures   \cite{endoCarbon1999}.    For   higher   concentration   of
substitutional boron atoms,  around 2.4\%, we find that  the two boron
atoms  lying  on  the  opposite  sites  of  the  same  hexagonal  ring
represents  the energetically  most stable  structure.   This geometry
represents    the   building   block    of   the    $C_6B$   structure
\cite{hachCarbon1999}.   The formation  of H  dimer like  structure on
graphene  has been  confirmed, as  well  as the  increase of  hydrogen
binding  energy nearby  the boron  substitutional sites.   However, in
contrast with the  previous studies, here we have  considered the most
likely  configuration  for  the  substitutional  boron  atoms  on  the
graphene surface.   We find that  the formation of H  dimer structure,
neighboring a single boron  substitutional site, is energetically less
favorable  compared  with two  isolated  H  monomers.  Meanwhile,  the
formation of H dimer structure is expected for higher concentration of
boron  atoms,  i.e.,  two  substitutional boron  atoms  occupying  the
opposite sites of  a hexagonal carbon ring, $B1$-$B2$.   In this case,
the binding  energy of H adatoms  is higher compared  with the binding
energy of an isolated H$_2$ molecule.

\section{Method of Calculations}
 
Our calculations were performed in the framework of the spin polarized
density  functional theory  (DFT)~\cite{kohn}, within  the generalized
gradient approximation due to Perdew, Burke, and Ernzerhof~\cite{PBE}.
The electron--ion  interaction was treated  by using norm--conserving,
{\it  ab  initio}, fully  separable  pseudopotentials \cite{KL}.   The
Kohn--Sham  wave   functions  were   expanded  in  a   combination  of
pseudoatomic numerical orbitals~\cite{sankey}.   Double zeta basis set
including polarization  functions (DZP)  was employed to  describe the
valence  electrons~\cite{dzp}.    The  self--consistent  total  charge
density was obtained by using the SIESTA code~\cite{siesta}.  Graphene
sheets were described within the supercell approach, by a single layer
of graphene with  84 atoms, separated by 15~\AA\  from their image.  A
mesh cutoff of 170~Ry was  used for the reciprocal--space expansion of
the total charge density, and  the Brillouin zone was sampled by using
up to 15  special {\bf k} points. We have  verified the convergence of
our results with respect to the  number and choice of the special {\bf
k} points.  All  atoms of graphene sheets were  fully relaxed within a
force convergence criterion of 20 meV/\AA.

\section{Results and Comments}

\subsection{Boron}

\begin{figure}[h]
 \includegraphics[width= 8cm]{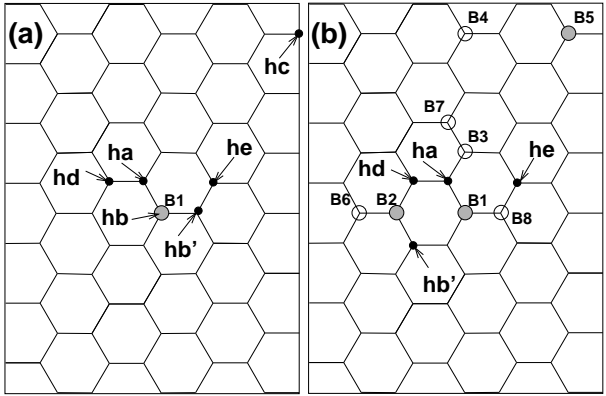}
 \caption{Structural models  of boron doped graphene sheet  for, (a) a
single substitutional  boron atom  ($B1$), and (b)  two substitutional
boron  atoms   per  unit   cell  ($B1-B2$),  corresponding   to  boron
concentrations of $\sim$1.2 and 2.4~\%, respectively.}
 \label{frame-new}
 \end{figure}

Initially  we examined  the  equilibrium geometry  and the  electronic
properties of boron doped  graphene sheet. For a single substitutional
boron atom,  $B1$ in Fig.~\ref{frame-new}(a), we obtained  a B--C bond
length (d$_{\rm  B-C}$) of 1.50~\AA,  while the nearest  neighbor C--C
bonds are  slightly compressed  (by $\sim$0.01~\AA) compared  with the
undoped  system.  The  structural  deformations of  graphene are  very
localized  around  the substitutional  boron  site.  Experimental  STM
measurements  indicate d$_{\rm  B-C}$ of  1.59~\AA\ \cite{endoJAP2001}
which is in agreement with our calculated results.

\begin{figure}[h]
 \includegraphics[width= 8cm]{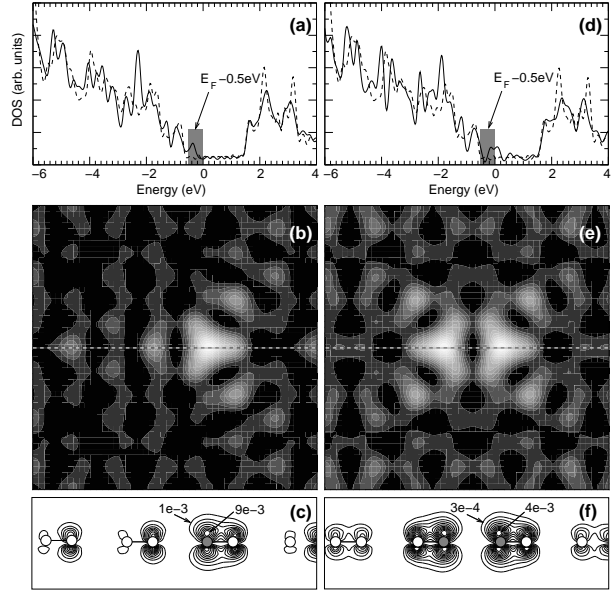}
 \caption{  Calculated  density  of  states (a)/(d),  STM  images  for
occupied states  within an energy  interval of 0.5~eV below  the Fermi
level  (b)/(e), and  electronic distribution  of the  highest occupied
states  ($E_F -  0.5~\rm  eV$)  along the  C--B  bonds (c)/(f)  (unit:
$e$/bohr$^3$), for the structural models $B1$/$B1$-$B2$.}
  \label{dosB1B2Orig}
 \end{figure}

Graphene sheets are metallic, as  indicated by density of states (DOS)
diagrams shown in Figs.~\ref{dosB1B2Orig}(a) and \ref{dosB1B2Orig}(d),
dashed lines.  Such a metallic character has been kept for boron doped
graphene (solid lines), however, the electronic density of states near
the valence  band maximum increases compared with  the pristine system
(shaded    regions).     Figure~\ref{dosB1B2Orig}(b)   presents    the
localization  of the electronic  states within  an energy  interval of
0.5~eV below the  Fermi level ($E_F - 0.5~\rm  eV$).  In this diagram,
the higher lying  occupied states are projected onto  a parallel plane
at  1.25~\AA\ from  the  graphene sheet.   Within the  Tersoff--Hamman
approach~\cite{tersoff},  this  projected   local  density  of  states
[Fig.~\ref{dosB1B2Orig}(b)]   corresponds  to   the   STM  images   of
boron--doped   graphene   sheet.   The   bright   protrusion  on   the
substitutional boron  (center) and  the nearest neighbor  carbon atoms
(edges) indicate  that the occupied  electronic states, within  $E_F -
0.5~\rm eV$,  are concentrated  around the boron  substitutional site.
Those bright spots  come from $\pi$ bonding states  of $2p_z$ orbitals
along the B--C bonds, see Fig.~\ref{dosB1B2Orig}(c).  The formation of
bright regions around the substitutional B atoms is in accordance with
the experimentally obtained STM pictures~\cite{endoJAP2001}.

Within  our supercell  approach,  a single  substitutional boron  atom
corresponds to a boron  concentration of $\sim$1.2~\%. Two boron atoms
per supercell  corresponds to  a boron concentration  of $\sim$2.4~\%.
In  this   case,  we  have  considered   several  plausible  $Bi$-$Bj$
configurations, indicated in  Fig.~\ref{frame-new}(b), for boron atoms
occupying substitutional  sites on  graphene sheet.  Our  total energy
results, summarized  in Table~\ref{energies}, reveal  the formation of
preferential    domains,    or    preferential   configurations    for
substitutional boron  atoms on graphene,  {\it viz.}: two  boron atoms
occupying the opposite sites of  the same hexagonal ring, $B1$-$B2$ in
Fig.~\ref{frame-new}(b),  represents  the  energetically  most  stable
configuration.  Whereas $B1$-$B8$ is the energetically least favorable
structure.  The  former geometry, $B1$-$B2$,  was proposed by  Hach et
al.  for ``boron  rich carbon structures''~\cite{hachCarbon1999}.  The
structural models $B1$-$B4$ and  $B1$-$B5$ are slightly less favorable
compared with  $B1$-$B2$. Since in  $B1$-$B4$ and $B1$-$B5$  the boron
atoms  are placed  far from  each other,  d(B--B) =  7.5  and 8.6~\AA,
respectively.   We can  infer that  at low  concentration  regime, the
substitutional boron atoms may spread out the graphene sheet. Whereas,
under B-rich  condition the  formation of $B1$-$B2$  becomes dominant,
giving rise  to $C_6B$ like  structures~\cite{hachCarbon1999}. Indeed,
$B1$-$B2$  represents  the  building  block  of  the  proposed  $C_6B$
structure.

\begin{table}
\caption{\label{energies}   Total  energy   differences,   $\Delta  E$
(eV/B atom), and B--B distances in \AA.}
\begin{ruledtabular}
\begin{tabular}{ccc}
Configuration    & $\Delta E$ & d(B--B)  \\
\hline
$B1$-$B2$        & 0.00 &  2.87      \\
$B1$-$B3$        & 0.17 &  2.53      \\
$B1$-$B4$        & 0.02 &  7.49          \\
$B1$-$B5$        & 0.01 & 8.64       \\
$B1$-$B6$        & 0.08 & 4.33       \\
$B1$-$B7$        & 0.12 & 3.86       \\
$B1$-$B8$        & 0.66 & 1.58       \\
\end{tabular}
\end{ruledtabular}
\end{table}

Figure~\ref{dosB1B2Orig}(d) (solid  line) presents our  calculated DOS
for  the structural model  $B1$-$B2$. We  observe that  the electronic
density of states increases  near valence band maximum (shaded region)
compared  with  the  undoped   system  (dashed  line).   The  occupied
electronic states  within $E_F -  0.5~\rm eV$ are mainly  localized on
the substitutional boron and neighboring the carbon atoms, giving rise
to  the  two bright  protrusions  shown in  Fig.~\ref{dosB1B2Orig}(e).
Figure~\ref{dosB1B2Orig}(f)  indicates   that  such  a  STM  picture
originates from the $\pi$ orbitals along the B--C bonds.

It  is  worth  to  pointing   out  that  the  unlikely  $B1$-$B3$  and
$B1$-$B7$~\cite{zhuJPhysChemB2006}                                  and
$B1$-$B8$~\cite{ferroJChemPhys2003}  arrangements ($\Delta E$  = 0.17,
0.12, and 0.66~eV/B atom,  respectively) were considered as substrates
for  hydrogen adsorption on  boron doped  graphene.  Thus,  we believe
that  further studies  are  necessary addressing  the  role played  by
substitutional boron impurities for hydrogen adsorption processes, now
considering  the  energetically  most  stable  configuration  for  the
substitutional boron atoms, $B1$-$B2$.
 
\subsection{Hydrogen}

Having  established  the energetically  most  probable structures  for
substitutional boron atoms on  graphene, {\it viz.}: structural models
$B1$          [Fig.~\ref{frame-new}(a)]          and         $B1$-$B2$
[Fig.~\ref{frame-new}(b)],  in this section  we examined  the hydrogen
adsorption processes on the clean and boron doped graphene sheets.

Firstly,  using the 84-atoms  unit cell  [Fig.~\ref{frame-new}(a)], we
calculate the  binding energy of  hydrogen adatoms on  clean graphene.
%%The hydrogen adsorption energy ($E^{\rm ads}$) can be written as:
%%$$
%%E^{\rm ads} = E[{\rm graphene + n_H H}] - 
%%              E[{\rm graphene}] - {\rm n_H}E[{\rm H}].
%%$$
%%$ E[{\rm  graphene}]$ and  $E[{\rm graphene +  n_H H}]$  represent the
%%total energies of clean  and hydrogen adsorbed graphene, respectively.
%%${\rm n_H}$ indicates the number  of hydrogen adatoms per surface unit
%%cell,  and $E[{\rm  H}]$ represents  the total  energy of  an isolated
%%hydrogen atom. 
For a single hydrogen adatom  we find a binding energy ($E^b_{H}$) of
0.98~eV/H-atom.  The  C--H equilibrium bond length  is 1.14~\AA, which
indicates the formation of a  C--H covalent bond, while the carbon
atom underneath the hydrogen adatom  moves upward by 0.5~\AA\ from the
flat graphene layer.  For  an isolated hydrogen molecule, we calculate
a  binding energy  of 4.34~eV/H$_2$-molecule  (2.17~eV/H-atom).  Thus,
with  respect  to  the  H$_2$  molecule, the  hydrogen  adsorption  on
graphene  is  an endothermic  process  by  1.19~eV/H-atom.  Our  total
energy and equilibrium geometry results are in agreement with previous
{\it ab initio} studies of  hydrogen adsorption on the graphene sheet,
for instance, $E^b_{H}$ of 0.7--0.8~eV/H-atom and C--H bond length of
1.13~\AA~~\cite{shaSUSC2002,duplockPRL2004,hornekaerPRL2006}.

In a very recent study, Hornek{\ae}r et al. observed that the presence
of a  hydrogen adatom on  graphene surface gives rise  to preferential
adsorption sites (nearby the already adsorbed hydrogen adatom) for the
subsequent hydrogen  sticking process~\cite{hornekaerPRL2006}.  Indeed
we find  that a  hydrogen adatom lying  on $ha$ increases  the binding
energy of the  next hydrogen adatom ($E^b_{HH}$) on  $hd$ or $he$ [cf.
Fig.~\ref{frame-new}(a)].   We calculate  $E^b_{HH}$  = 2.03~eV/H-atom
for   both  hydrogen  configurations,   $ha/hd$  and   $ha/he$,  while
Hornek{\ae}r   et  al.    obtained   binding  energies   of  1.9   and
2.1~eV/H-atom,   respectively.   Those   results  indicate   that  the
formation  of $ha/hd$ or  $ha/he$ hydrogen  dimer like  structures are
energetically more favorable compared with two isolated $ha$ monomers,
that  is,  $ha$+$ha$  $\rightarrow$  $ha/hd$, $ha/he$  are  exothermic
processes.   Our  findings are  in  accordance  with previous  studies
related to the formation of hydrogen clusters on graphene surface.  In
addition, our calculated binding  energies $E^b_{HH}$ (for $ha/hd$ and
$ha/he$) are comparable  with the binding energy of  an isolated H$_2$
molecule  (2.17~eV/H-atom).   Suggesting  that  presence of  a  single
hydrogen adatom ($ha$) on the  graphene sheet, somewhat  promote the
H$_2$ dissociation nearby $ha$.

\begin{table} 
\caption{\label{energies-H} Binding energies ($E^b$) of hydrogen
adatoms on clean and boron doped graphene surface in eV/H-atom.}
\begin{ruledtabular}
\begin{tabular}{cccccccc}
\multicolumn{1}{c}{Model}   &
\multicolumn{2}{c}{$E^b_{H}$} &
\multicolumn{5}{c}{$E^b_{HH}$} \\
         & $ha$ &$ hb$  &$ha/hb$  &$ha/hb'$  &$ha/hc$  &$ha/hd$  &$ha/he$ \\
\cline{1-1} \cline{2-3} \cline{4-8}
clean    & 0.98 & --    &  $-$    & $-$      &  $-$    & 2.03    & 2.03 \\
$B1$     & 1.89 & 1.51  & 1.36    & 1.51     & 1.00    & 1.67    & 1.25 \\
$B1$-$B2$& 2.13 & $-$   &   $-$   & 1.73     &  $-$    & 2.37    & 1.64 \\  
\end{tabular}
\end{ruledtabular}
\end{table}

\begin{figure}[h]
 \includegraphics[width= 8cm]{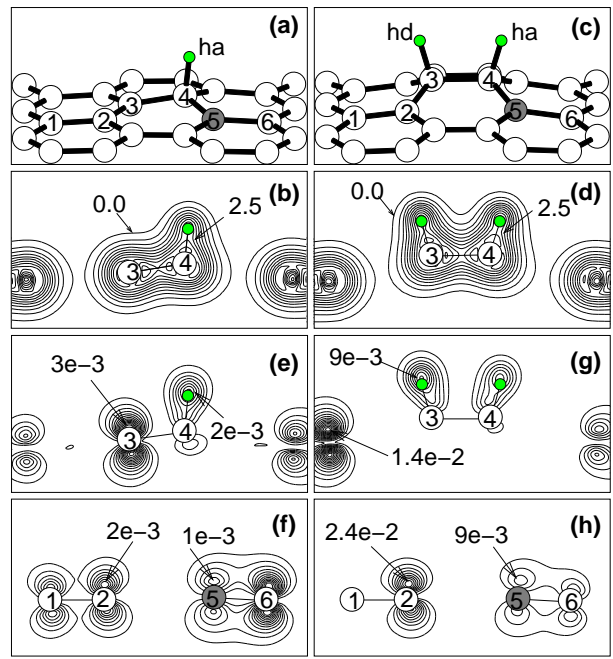}
 \caption{Equilibrium  geometry and total  charge density  of hydrogen
adatoms on the $B1$ graphene sheet, (a)--(b) $ha$ and (c)--(d) $ha/hd$
models.  Partial electronic charge density within $E_F-0.5~\rm eV$ for
(e)--(f) $ha$ and (g)--(h) $ha/hd$ models. Unit, $e$/bohr$^3$.}
 \label{B1H1H2}
 \end{figure}

The binding  energy of a single hydrogen  adatom, $E^b_{H}$, increases
upon  the  presence of  substitutional  boron  atoms  on the  graphene
surface.   Our   results  of   binding  energies  are   summarized  in
Table~\ref{energies-H}.   For the  $ha/B1$ configuration,  depicted in
Fig.~\ref{B1H1H2}(a), we  find $E^b_{H}$ =  1.89~eV/H-atom, whereas on
the  clean graphene  surface we  have $E^b_{H}$  =  0.98~eV/H-atom. So
that, we  can infer that presence  of a substitutional  boron atom, in
the vicinity  of the hydrogen  adsorption site, strengthened  the C--H
bond by  0.91~eV.  Meanwhile,  the hydrogen adsorption  on top  of the
substitutional boron atom, $hb/B1$, is energetically less favorable by
0.38~eV/H-atom  compared with $ha/B1$.   At the  equilibrium geometry,
the  C--H bond  length is  1.15~\AA\ and  the C  atom  (underneath the
hydrogen  adatom) moves  upward  by 0.4~\AA.   The  formation of  C--H
covalent     bond     is     depicted     in     Fig.~\ref{B1H1H2}(b).
Figures~\ref{B1H1H2}(e)  and   \ref{B1H1H2}(f)  present  the  occupied
electronic states along the  C--H and B--C bonds, respectively, within
$E_F -  0.5~\rm eV$  . Those electronic  states are composed  by $\pi$
orbitals of  carbon and (substitutional)  boron atoms on  the graphene
sheet,  and   $sp_z$  $\sigma$  hybridization  along   the  C--H  bond
[Fig.~\ref{B1H1H2}(e)].   Comparing the electronic  densities depicted
in  Figs.~\ref{dosB1B2Orig}(c) and  \ref{B1H1H2}(f), $B1$  and $ha/B1$
structures, respectively, we can infer that the hydrogen adsorption on
$ha$ reduces  the electronic density of occupied  $\pi$ orbitals along
the B--C bonds.

Keeping  the   $B1$  configuration  adsorbed  by   a  hydrogen  adatom
($ha/B1$),  the subsequent  hydrogen adsorption  has been  examined by
considering  a  number of  plausible  configuration  for two  hydrogen
adatoms on  $B1$ graphene sheet.   Our binding energy results  for the
hydrogen adatom  on the  $ha/B1$ graphene sheet,  $E^b_{HH}$, indicate
that $ha/hc$ represents  the energetically least stable configuration.
We find $E^b_{HH}$ =  1.00~eV/H-atom, nearly the same hydrogen binding
energy on the clean graphene, $E^b_{H}$ = 0.98~eV/H-atom.  This result
is somewhat expected, since $hc$  is far from the substitutional boron
site,  $B1$. Figure~\ref{B1H1H2}(c)  presents  the energetically  most
stable configuration, $ha/hd$,  $E^b_{HH}$ = 1.67~eV/H-atom.  The C--H
bond  length is  equal  to  1.14~\AA, and  its  covalent character  is
depicted in Fig.~\ref{B1H1H2}(d).  Similar  to the $ha/B1$ system, the
electronic states  within $E_F  - 0.5~\rm eV$  are composed  by $sp_z$
$\sigma$  hybridization along the  C--H bonds,  and $\pi$  orbitals of
carbon  and boron  atoms,  Figs.~\ref{B1H1H2}(g) and  \ref{B1H1H2}(h),
respectively.   However, in this  case, the  adsorption of  the second
hydrogen adatom ($hd$) is not favored by the presence of the first one
($ha$), $E^b_{HH}$  is lower than  $E^b_{H}$.  Consequently, $ha$+$ha$
$\rightarrow$ $ha/hd$ is an endothermic process, thus, indicating that
the formation  of hydrogen  clusters is not  expected to  occur nearby
single  boron  substitutional sites.   Thus,  for  higher coverage  of
hydrogen adatoms,  we can  infer that hydrogen  clusters on  the clean
region  graphene surface,  i.e.  far  from boron  substitutional sites
(with $E^b_{HH}\approx 2.0$~eV/H-atom)  are more likely than hydrogen
adatoms   neighboring   an    isolated   boron   substitutional   site
($E^b_{HH}\approx 1.7$~eV/H-atom).

\begin{figure}[h]
 \includegraphics[width= 8cm]{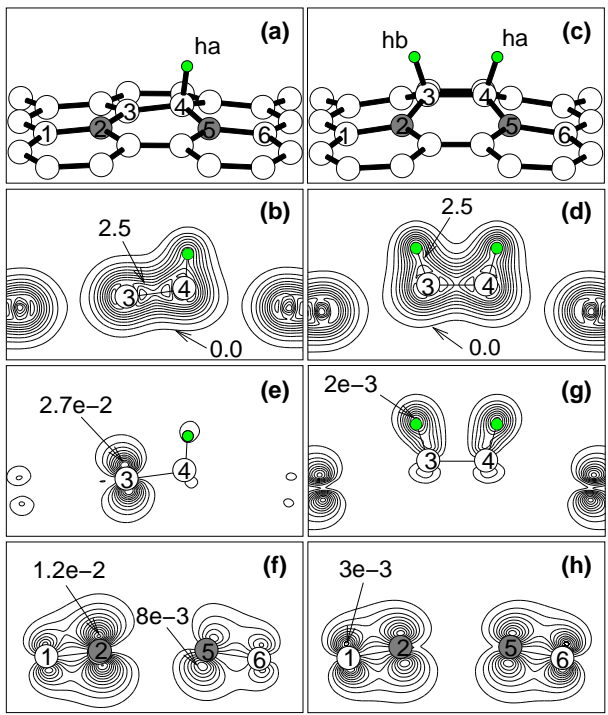}
 \caption{Equilibrium  geometry and total  charge density  of hydrogen
adatoms on  the $B1$-$B2$ graphene  sheet, (a)--(b) $ha$  and (c)--(d)
$ha/hd$ models.  Partial electronic charge density within $E_F-0.5~\rm
eV$   for   (e)--(f)  $ha$   and   (g)--(h)   $ha/hb$  models.   Unit,
$e$/bohr$^3$.}
 \label{B2H1H2}
 \end{figure}

Increasing  the concentration of  substitutional boron  atoms, forming
the  energetically   most  probable  configuration,   i.e.   $B1$-$B2$
[Fig.~\ref{frame-new}(b)],  the  binding  energy of  hydrogen  adatoms
increases  compared with  the  $B1$ graphene  surface.   For a  single
hydrogen  adatom,  $ha$/$B1$-$B2$  in  Fig.~\ref{B2H1H2}(a),  we  find
$E^b_{H}$  equal  to  2.13~eV/H-atom,  which is  comparable  with  the
calculated H$_2$ binding  energy (2.17~eV/H-atom).  At the equilibrium
geometry, the C--H bond length  is 1.15~\AA, and the formation of C--H
covalent     bond     is     depicted     in     Fig.~\ref{B2H1H2}(b).
Figure~\ref{B2H1H2}(e)  shows that the  $sp_z$ $\sigma$  orbitals does
not lie within  $E_F - 0.5~\rm eV$.  Whereas the  $p_z$ orbital of the
carbon atom C(3) strongly  contributes to the electronic states within
this energy interval.  The former $\sigma$ bonding states, composed by
hydrogen  $s$  and carbon  $p_z$  orbitals,  are  resonant within  the
valence  band   of  the   graphene  sheet.   In   addition,  comparing
\ref{dosB1B2Orig}(f)  and \ref{B2H1H2}(f)  we  verify that  electronic
density   of   $\pi$  orbitals   along   the   B--C  bonds   increases
(asymmetrically) upon hydrogen adsorption.

Different from  $ha/hd$ on the  $B1$ graphene sheet, the  formation of
$ha/hd$     dimer    like     structure    on     $B1$-$B2$,    namely
$ha$+$ha$~$\rightarrow$~$ha/hd$,   is   an   exothermic   process   by
0.24~eV/H-atom.   Furthermore, $E^b_{HH}$ is  higher than  the binding
energy of  H$_2$ molecule,  that is, the  formation of $ha/hd$  on the
boron  doped $B1$-$B2$ graphene  surface is  an exothermic  process by
0.20~eV/H-atom compared with free H$_2$ molecules.  Thus, we can infer
that  the  formation of  hydrogen  clusters  can  be improved  by  the
increase  of  substitutional  boron  atoms  forming  the  $C_6B$  like
structure  on the  graphene surface.   Figure~\ref{B2H1H2}(d) presents
the total  charge density  along the C--H  covalent bonds.  The carbon
atoms bonded  to hydrogen adatoms exhibit  $sp^3$ like hybridizations,
where C(3) and C(4) are  displaced upward by 0.94~\AA\ with respect to
the pristine  graphene sheet.  The C(3)--C(4)  equilibrium bond length
(1.54~\AA) is stretched by 0.1~\AA\ compared with the C--C bond length
of  undoped  systems.  The  localization of  the  occupied  electronic
states,    within   $E_F    -   0.5~\rm    eV$,   are    depicted   in
Figs.~\ref{B2H1H2}(g) and  \ref{B2H1H2}(h).  In the  latter diagram we
find that the electronic density  of the $\pi$ orbitals along the B--C
bonds  is  reduced  compared   with  the  $\pi$  orbitals  of  $ha/B1$
[Fig.~\ref{B2H1H2}(f)].   Figure~\ref{B2H1H2}(g)  depicts the  $sp_z$
$\sigma$ orbitals along  the C--H bonds.  In a  STM measurement, those
$sp_z$ states will  give rise to bright spots  on the hydrogen adatoms
lying on  the $B1$-$B2$ boron  doped graphene sheets.   Indeed, recent
STM pictures,  supported by {\it ab initio}  simulations, indicate the
formation  of bright  protrusions attributed  to the  hydrogen adatoms
(forming    a    dimer    like    structure)   on    clean    graphene
sheets~\cite{hornekaerPRL2006-ii}. On the  other hand, quite different
STM pictures are  expected for $ha$/$B1$-$B2$. Figure~\ref{B2H1H2}(e)
indicates that,  for STM bias  voltage of $\sim$0.5~V below  the Fermi
level,  the  tunneling  current  from  the hydrogen  adatoms  will  be
negligible compared with the one from neighbor carbon and boron atoms.
Thus, in this  case, the hydrogen adatom site  will appear darker than
the neighboring carbon and boron atoms.

\section{Conclusions}

In  summary,  we  have  performed  an {\it  ab  initio}  total  energy
investigation of  boron doped  graphene sheets, and  their interaction
with  hydrogen adatoms. For  a single  substitutional boron  atom, the
equilibrium geometry of graphene  sheet is weakly perturbed, while the
electronic density  of states near  the valence band maximum  has been
increased.  Our simulated STM  images indicate the formation of bright
(triangular) spots  on the boron substitutional site  (center) and the
nearest neighbor carbon atoms  (edges).  For two boron atoms occupying
the  substitutional sites on  the graphene  sheet, corresponding  to a
boron concentration of $\sim$2.4\%, we  find that boron atoms lying on
the opposite  sites of the  same hexagonal ring  ($B1$-$B2$ structure)
represents the energetically  most stable configuration.  Furthermore,
the hydrogen adsorption on graphene has been examined as a function of
the concentration  of boron atoms.  For the  pristine graphene system,
we find  that the binding energy  of hydrogen adatoms  forming H dimer
like  configurations ($ha/hd$  or  $ha/he$) is  higher by  1~eV/H-atom
compared with the binding energy of two isolated H monomers ($E^b_H$ =
0.98~eV/H-atom).    This   is    in   accordance   with   the   recent
experimental/theoretical  study  of the  formation  of  H clusters  on
graphene    sheets~\cite{hornekaerPRL2006-ii,hornekaerPRL2006}.    The
presence of a single  substitutional boron atom increases the hydrogen
binding energy  by 0.9~eV/H-atom  compared with the  one on  the clean
graphene sheet. In  this case, the hydrogen adatom lies  on top of the
carbon  atom nearest neighbor  to the  substitutional B  site ($ha/B1$
configuration).   However,  different from  the  pristine system,  the
adsorption of  two hydrogen adatoms close  to each other,  forming a H
dimer like  structure, is  energetically less favorable  compared with
two isolated H monomers.  This  result indicates that the formation of
hydrogen clusters on the graphene  sheet is somewhat suppressed by the
presence of  a single substitutional  boron atom.  Finally,  for boron
concentration  of $\sim$2.4\%  we find  that the  binding energy  of H
dimer structure  ($E^b_{HH}$) is  higher than the  binding energy  of an
isolated H$_2$ molecule, where the  hydrogen adatoms lie on the carbon
atoms neighboring the substitutional  boron sites. In this case, boron
atoms give rise to  preferential adsorption sites for hydrogen adatoms
on  graphene sheets,  and thus,  improving the  formation  of hydrogen
clusters on graphene sheets. Those results indicate that the formation
(or  not)  of H  clusters  on  graphene sheets  can  be  tuned by  the
concentration of substitutional boron atoms.

\begin{center}
  {\large\bf Acknowledgements}
\end{center}

The authors acknowledge financial support from the Brazilian agencies
CNPq, FAPEMIG, and  FAPESP, and the computational facilities of the
Centro Nacional de Processamento de Alto Desempenho/CENAPAD-Campinas.

%%%%References from RHMiwa.bib,
%%%%
\bibliography{/home/rohiroki/Trab/RHMiwa}
%%%%

\end{document}